\documentclass[11pt,a4paper]{article}
\usepackage[T1]{fontenc}
\usepackage[utf8]{inputenc}
\usepackage{amsmath,amssymb,bm}
\usepackage{booktabs}
\usepackage{braket}
\usepackage[margin=2.7cm]{geometry}
\usepackage{setspace}
\usepackage[colorlinks=true,linkcolor=blue,citecolor=blue]{hyperref}
\hypersetup{pdftitle={Kramers pseudospin and the quantum number proposed for the many-electron Dirac--Coulomb Hamiltonian},pdfauthor={Shadan Ghassemi Tabrizi}}

\newcommand{\ad}[1]{a^{\dagger}_{#1}}
\newcommand{\an}[1]{a^{\phantom{\dagger}}_{#1}}
\newcommand{\bp}{\bar{p}}
\newcommand{\Fnorm}[1]{\lVert #1 \rVert_{F}}
\newcommand{\up}{\uparrow}
\newcommand{\dn}{\downarrow}

\title{Kramers pseudospin and the quantum number proposed for the many-electron Dirac--Coulomb Hamiltonian}
\author{Shadan Ghassemi Tabrizi\textsuperscript{1,\,2,}\thanks{s.ghassemi-tabrizi@hzdr.de}\\[4pt]
\footnotesize\textsuperscript{1}\,Computational System Sciences, Technische Universit\"at Dresden, 01187 Dresden, Germany\\[1pt]
\footnotesize\textsuperscript{2}\,Center for Advanced Systems Understanding (CASUS), Am Untermarkt 20, 02826 G\"orlitz, Germany}
\date{}

\begin{document}
\maketitle

\begin{abstract}
\noindent
The square of a sum of single-electron time reversals has been proposed
as a conserved quantity with an integer eigenvalue spectrum for the
many-electron Dirac--Coulomb Hamiltonian [Phys.~Rev.~A \textbf{94}, 052104 (2016)], and its expectation value is in use as a diagnostic of
Kramers contamination. This work identifies the operator behind the construction, a component of a pseudospin attached to the chosen Kramers pairs, whose angular-momentum algebra reproduces the reported spectrum and yields the eigenvectors in closed form. The label $k$ entering the proposed quantum number $-k^{2}$ is twice the magnitude of a pseudospin projection onto an axis fixed by the basis and changes when the Kramers pairs are rephased. The commutation with the Hamiltonian is unfounded, and the proposed quantum number does not stand.
\end{abstract}

\section{Introduction}
\label{sec:intro}

In nonrelativistic and scalar-relativistic electronic-structure theory, the
levels of a many-electron system can be classified by total spin and its
projection, because the Hamiltonian commutes with $\mathbf{S}^2$ and $S_z$. When
spin--orbit coupling is treated variationally, as in four-component theory based on the Dirac--Coulomb Hamiltonian $H^{\mathrm{DC}}$, this classification is lost. Apart from point-group symmetry, the remaining symmetry (for fixed nuclei and in the absence of external magnetic fields) is time reversal~\cite{Wigner1932,Sakurai}, whose consequences are Kramers
degeneracy~\cite{Kramers1930} and certain constraints on matrix
elements~\cite{AbragamBleaney1970}.
Relativistic many-electron methods typically work in bases whose spinors are
paired with their time
reverses~\cite{Aucar1995,Jensen1996,SaueJensen1999}, but this pairing, unlike
spin adaptation in the nonrelativistic case, corresponds to no quantum number
of the Hamiltonian.

Bu\v{c}insk\'y \emph{et al.}~\cite{Bucinsky2015,Bucinsky2016} and
Komorovsk\'y, Repisk\'y and Bu\v{c}insk\'y~\cite{Komorovsky2016} have
proposed a conserved quantity of $H^{\mathrm{DC}}$ built from time reversal. They form a sum $K_+$ of the single-electron time reversals of
the $N$ electrons and represent its square $K_+^2$ in a basis of
Kramers-restricted Slater determinants, which yields a real symmetric matrix
with eigenvalues $-k^2$, where $k$ is a nonnegative integer of the parity of
$N$, bounded by the number of unpaired electrons.
Reference~\cite{Komorovsky2016} asserts that this matrix commutes with the
Hamiltonian represented in the same basis, concludes that $-k^2$ is a
quantum number, and proposes the eigenvectors as relativistic configuration
state functions. An algorithm generating the $K_+^2$ matrix for arbitrary numbers of open shells is supplied in Ref.~\cite{Gall2018}, and the expectation value $\langle K_+^2\rangle$ is in use as a diagnostic of Kramers
contamination~\cite{KasperLi2020}. The physical content of the additive construction has remained open. Gall \emph{et al.}\ name it as what
``remains as a challenge''~\cite{Gall2018}.

The present work settles the interpretation. It does not dispute the published matrices, spectra, or eigenvectors, but identifies their operator content and asks whether that content supports the proposed quantum number. The matrix of
Refs.~\cite{Bucinsky2015,Bucinsky2016,Komorovsky2016}, an array of inner
products of well-defined states, is the matrix of a Cartesian component of a pseudospin attached to the chosen Kramers pairs. Its angular-momentum algebra reproduces the reported spectrum, and the eigenvectors follow in closed form. The label $k$ is accordingly $2\lvert m_y\rvert$, twice the magnitude of the pseudospin projection $m_y$ onto an axis fixed by the chosen basis of Kramers pairs, and it
changes when the pairs are rephased. A sum of single-electron
time reversals, taken literally, defines no operator on the
many-electron Hilbert space, because time reversal is antilinear and an
antilinear operation cannot be confined to one electron of a many-electron state. Every well-defined antilinear realization of the proposed sum contains a single global conjugation, and the realization attached to the chosen Kramers basis reproduces the published matrix. The asserted commutation with the Hamiltonian does not follow from time-reversal invariance, and the ordinary Coulomb exchange between two Kramers pairs, itself time-reversal invariant, violates it. That construction, and with it the commutation, depends on the chosen pairing (Section~\ref{sec:label}). In the spin-free case with a common spin axis the commutation does hold, as
a consequence of ordinary spin symmetry, the setting of the published comparisons of the eigenfunctions with those of $\mathbf{S}^2$~\cite{Komorovsky2016}. Averaged over the pair axes of a
fixed Kramers pairing, the additive construction retains the number of
unpaired electrons, the content of its diagnostic use.
Section~\ref{sec:sum} examines the proposed sum, Sections~\ref{sec:operator} and~\ref{sec:label} the operator behind the $K_+$ matrix and the label it carries,
and Sections~\ref{sec:claim} and~\ref{sec:derivation} the asserted commutation and the derivation published for it; Section~\ref{sec:valid} collects what remains valid.

\section{The proposed sum of single-electron time reversals}
\label{sec:sum}

For a single electron, time reversal is the antiunitary operator
\begin{equation}
  \theta=-\mathrm{i}\Sigma_yK_0 ,
  \label{eq:theta}
\end{equation}
a factorization into a rotation times a conjugation, taken relative to the
standard representation. The operator $K_0$ of complex conjugation is
defined only relative to a basis. It leaves the chosen basis kets unchanged
and conjugates the expansion coefficients of every state, so that
conjugating the coefficients of a different expansion defines a different
map. In Eq.~\eqref{eq:theta} it refers to the standard representation, in
which $\sigma_y$ is the only imaginary Pauli matrix. The factor $-\mathrm{i}\Sigma_y$ is unitary, the spinor representation of a rotation by $180^\circ$ about the $y$ axis, $\Sigma_y$ being the four-row counterpart of $\sigma_y$. The composition satisfies $\theta^2=-1$, and with
antiunitarity it follows that $\braket{\psi|\theta\psi}=0$ for every spinor
$\psi$~\cite{Sakurai}. An orthonormal basis of the $2m$-dimensional spinor
space can therefore be arranged in Kramers pairs,
\begin{equation}
  B=\{\,e_p,\ \bar e_p\equiv\theta e_p\,\}_{p=1}^{m} ,
  \label{eq:frame}
\end{equation}
the two partners of a pair conventionally distinguished as unbarred and
barred. A basis of this form will be called a Kramers basis.

For $N$ electrons, the many-electron time-reversal operator $\Theta$ (denoted $\mathcal{K}$ in Ref.~\cite{Komorovsky2016}) applies $\theta$ to every
occupied spinor of a determinant,
\begin{equation}
  \Theta\,(v_1\wedge\cdots\wedge v_N)=\theta v_1\wedge\cdots\wedge\theta v_N ,
  \label{eq:Theta}
\end{equation}
where $v_1\wedge\cdots\wedge v_N$ is the Slater determinant with occupied
spinors $v_1,\dots,v_N$. This map is antiunitary and well defined. A scalar factor moved from one occupied spinor to another is conjugated either way, so both expressions of the same state receive the same image. Its square is the scalar
\begin{equation}
  \Theta^{2}=(-1)^{N} .
  \label{eq:Thetasq}
\end{equation}
The consequences of the invariance of the Hamiltonian under $\Theta$ are Kramers degeneracy for odd $N$ and constraints on matrix elements.

An action of $\theta$ on one electron of an $N$-electron state, by contrast,
cannot be defined. Consider two electrons and, since antisymmetrization
plays no role in the argument, the product state $v_1\otimes v_2$ in which
electron 1 occupies $v_1$ and electron 2 occupies $v_2$. A scalar factor
belongs to the state as a whole,
\begin{equation}
  (c\,v_1)\otimes v_2=v_1\otimes(c\,v_2) ,
  \label{eq:samevector}
\end{equation}
as an identity of vectors. A map acting as $\theta$ on electron 1 and as the
identity on electron 2 would assign to the two sides of
Eq.~\eqref{eq:samevector} the images
\begin{equation}
  (c\,v_1)\otimes v_2 \ \longmapsto\ c^{*}\,(\theta v_1)\otimes v_2 ,
  \label{eq:nogo1}
\end{equation}
\begin{equation}
  v_1\otimes(c\,v_2) \ \longmapsto\ c\,(\theta v_1)\otimes v_2 ,
  \label{eq:nogo2}
\end{equation}
which differ for every nonreal $c$. One vector would receive two images.
The same identification of scalars holds in the antisymmetrized space,
$(c\,v_1)\wedge v_2=v_1\wedge(c\,v_2)$, and the antisymmetrizer is linear, so antisymmetrization does not remove the obstruction. Only the
antilinearity of $\theta$ entered this argument. Consequently, no operator of the $N$-electron space acts as $\theta$ on one electron and trivially on the others. The argument does not touch the product form in which Ref.~\cite{Komorovsky2016} writes the many-electron time reversal [its Eq.~(12)]. Uniform antilinearity is consistent where mixed linearity and antilinearity is not: $\theta\otimes\theta$ conjugates a scalar whichever factor carries it, so both sides of Eq.~\eqref{eq:samevector} receive the image $c^{*}(\theta v_1)\otimes\theta v_2$. A product form therefore has a well-defined referent, the map \eqref{eq:Theta}, whereas a sum requires its summands to be operators.

References~\cite{Bucinsky2015,Bucinsky2016,Komorovsky2016} seek a
label-carrying operator in a sum of the single-electron time reversals, which Ref.~\cite{Komorovsky2016} writes as
\begin{equation}
  \mathcal{K}_+=\sum_{i=1}^{N}K_i ,
  \label{eq:Kplus}
\end{equation}
written $K_+$ in what follows,
with
\begin{equation}
  K_i=-\mathrm{i}\,\Sigma_{y,i}\,K_{0,i}
  \label{eq:Ki}
\end{equation}
[Eqs.~(26) and (3) of that work]. For a single electron, $K_i$ is the
operator $\theta$ of Eq.~\eqref{eq:theta}. By the result just proved, no
operator of the $N$-electron space acts in this way on electron $i$ alone,
so Eqs.~\eqref{eq:Kplus} and \eqref{eq:Ki} are quoted here as the notation of Ref.~\cite{Komorovsky2016}, not as definitions of electron-wise operators.

The expression becomes well defined when the conjugation is read globally. If $K_0$ is understood as a single operation applied to the entire $N$-electron state in the standard representation, the index on $K_{0,i}$ merely repeating the electron label of $\Sigma_{y,i}$, then
$K_+=\bigl(\sum_i-\mathrm{i}\Sigma_{y,i}\bigr)K_0$ is well defined and
antilinear, because one global conjugation of a many-electron state is well
defined. Its square is $-4(S_y^{\Sigma})^2$, where $S_y^{\Sigma}$ is the $y$
component of the many-electron spin built from the matrices $\Sigma_y/2$.
The operator identified in Section~\ref{sec:operator} has the same shape, $-4$ times the square of a $y$ component; the two readings differ only in the spin to which the component belongs, that of the standard representation here and the pseudospin of the chosen pairing there. This operator has a spectrum bounded by the electron number $N$ rather than by the number of unpaired electrons, and a single closed Kramers pair already separates it from the matrix of Ref.~\cite{Komorovsky2016}. Take a two-component space with real spatial orbitals $\varphi_1,\varphi_2$ and set $\zeta=\tfrac{1}{\sqrt2}\bigl(\varphi_1+\mathrm{i}\varphi_2\bigr)$, so that $\braket{\zeta|\zeta^{*}}=0$. The determinant built from the single closed pair
\begin{equation}
  e=\zeta\,\alpha,\qquad
  \bar e=\theta e=\zeta^{*}\beta
  \label{eq:closedwitness}
\end{equation}
occupies two orthogonal spatial orbitals with opposite spins, so that $\langle \mathbf{S}^{2}\rangle=1$, $\langle S_z^{2}\rangle=0$ and $\langle S_y^{2}\rangle=\tfrac12$. This reading therefore returns $-4\langle (S_y^{\Sigma})^{2}\rangle=-2$ on this state, whereas the $K_+^2$ matrix returns $-N_O=0$: closed Kramers pairs contribute nothing to it [Appendix~G of that work], and its diagonal element is the value recorded in its Eq.~(80). The operator $-4(S_y^{\Sigma})^{2}$ is not the operator whose properties are reported in Ref.~\cite{Komorovsky2016}.

The conjugation can equally be referred to any other basis, each choice defining a different operator. No reading of the sum, however, is needed for the numbers that Ref.~\cite{Komorovsky2016} actually computes. Those numbers are the inner products
$K_{qp}=\braket{\phi_q|\theta\phi_p}$ of basis spinors with time-reversed
basis spinors, each well defined. Their array over the $2m$ spinors of a Kramers basis, inserted into the standard single-particle formula of second quantization, Eq.~\eqref{eq:secondq} below, defines a matrix on the space of Kramers-restricted Slater determinants -- the $K_+$ matrix, in what follows -- and its square, the $K_+^2$ matrix; both exist whether or not the sum $K_+$ of Eq.~\eqref{eq:Kplus} does. Every
quantitative statement of
Refs.~\cite{Bucinsky2015,Bucinsky2016,Komorovsky2016} about the construction -- the spectrum, the eigenfunctions, the commutation with the Hamiltonian -- is well defined only as a statement about matrices on the space spanned by the Kramers-restricted Slater determinants. Table~\ref{tab:objects} collects the objects of the analysis, including those of the next section.

\begin{table}[t]
\centering
\caption{The objects of the analysis. $K_+$ is the notation of Ref.~\cite{Komorovsky2016}; $B$ is the chosen Kramers basis \eqref{eq:frame}.}
\label{tab:objects}
\begin{tabular}{llp{7.2cm}}
\toprule
Object & Character & Role \\
\midrule
$K_+=\sum_iK_i$ & formal sum & no electron-wise operator behind it (Sec.~\ref{sec:sum}) \\
$\theta=-\mathrm{i}\Sigma_yK_0$ & antiunitary & single-electron time reversal \eqref{eq:theta} \\
$\Theta$ & antiunitary & $N$-electron time reversal \eqref{eq:Theta}, $\Theta^2=(-1)^N$ \\
$K_0$, $C_B$, $\mathcal{C}$ & antilinear & conjugations relative to the standard representation, to $B$, and to the determinant basis \\
$j_B$ & linear & pair rotation of $B$, $e_p\mapsto\bar e_p$, $\bar e_p\mapsto-e_p$; $\theta=j_BC_B$ (Sec.~\ref{sec:operator}) \\
$A_B=(j_B)_+=-2\mathrm{i}T_y$ & linear, antihermitian & operator behind the $K_+$ matrix (Sec.~\ref{sec:operator}) \\
$L_B=A_B\,\mathcal{C}$ & antilinear & realization of the sum attached to $B$ (Sec.~\ref{sec:operator}) \\
$U_\Theta=\exp\bigl(\tfrac{\pi}{2}A_B\bigr)$ & unitary & unitary factor of $\Theta=U_\Theta\,\mathcal{C}$ (Sec.~\ref{sec:operator}) \\
\bottomrule
\end{tabular}
\end{table}

\section{The Kramers pseudospin behind the \texorpdfstring{$K_+$}{K+} matrix}
\label{sec:operator}

From the summands $K_i$, Ref.~\cite{Komorovsky2016} derives its Eq.~(34) and records, for Kramers-restricted determinants, its Eq.~(35) -- three statements about the eigenvalues of the $K_+^2$ matrix: the
eigenvalues are $-k^{2}$ with $k$ a nonnegative integer, $k$ has the same
parity as $N$, and over Kramers-restricted Slater determinants $k$ is
bounded by the number $N_O$ of unpaired electrons. The manipulations employed there, $K_i^2=-1$ and $[K_i,K_j]=0$, treat
the summands as electron-wise operators, which by Section~\ref{sec:sum}
they are not. The statements are nevertheless correct. The $K_+$ matrix is the matrix of a well-defined linear operator, identified in Eq.~\eqref{eq:main} below, and under the antilinear realization of Eq.~\eqref{eq:antilinear} below the manipulations themselves become exact.

\subsection{Identification of the operator}

Equation~\eqref{eq:theta} factorizes time reversal relative to the standard representation. The same can be done relative to the Kramers basis $B$. Let $C_B$ denote the antilinear map that leaves the basis
kets $e_p,\bar e_p$ unchanged and conjugates expansion coefficients,
\begin{equation}
  C_B\sum_p\bigl(c_pe_p+c_{\bp}\bar e_p\bigr)
  =\sum_p\bigl(c_p^{*}e_p+c_{\bp}^{*}\bar e_p\bigr) .
  \label{eq:CB}
\end{equation}
$C_B$ is the analogue, for the basis $B$, of what $K_0$ is for the standard
representation, and like $K_0$ it satisfies $C_B^{2}=1$. Composing time
reversal with this conjugation,
\begin{equation}
  j_B\equiv\theta\,C_B ,
  \label{eq:split}
\end{equation}
produces a linear operator $j_B$, the composition of two antilinear
maps, and recovers the factorized form $\theta=j_B\,C_B$. Since $C_B$
leaves the basis kets unchanged, $j_B$ acts on them exactly as $\theta$
does,
\begin{equation}
  j_B\,e_p=\bar e_p ,
  \label{eq:jB}
\end{equation}
\begin{equation}
  j_B\,\bar e_p=\theta^{2}e_p=-e_p .
  \label{eq:jBbar}
\end{equation}
Consequently $j_B^{2}=-1$, and the matrix representing $j_B$ in the basis
$B$ is real and antisymmetric. The unitary factor of this splitting is
again a rotation by $180^\circ$; for the two partners of a Kramers pair,
$j_B$ plays the role that $-\mathrm{i}\sigma_y$ plays for the two spin
states of an electron, and the axis of the rotation is attached to the
basis. The two factorizations coincide only when the vectors of $B$ are
real in the standard representation, since $C_B$ and $K_0$ then agree. A
two-component example shows that they differ in general. For
$e=(1,\mathrm{i})^{\mathrm{T}}/\sqrt2$,
\begin{equation}
  \theta e = \tfrac{1}{\sqrt2}\,(\mathrm{i},\,1)^{\mathrm{T}} ,
  \label{eq:ex1}
\end{equation}
\begin{equation}
  -\mathrm{i}\sigma_y\,e = \tfrac{1}{\sqrt2}\,(-\mathrm{i},\,1)^{\mathrm{T}} ,
  \label{eq:ex2}
\end{equation}
and only $\theta e$ is orthogonal to $e$, as a Kramers partner must
be.

For a linear single-electron operator $u$, the many-electron sum
$u_+=\sum_{i=1}^{N}u_i$ takes the second-quantized form
\begin{equation}
  u_+=\sum_{qp}u_{qp}\,\ad{q}\an{p} ,
  \label{eq:secondq}
\end{equation}
with $u_{qp}=\braket{\phi_q|u|\phi_p}$, for any orthonormal single-particle
basis $\{\phi_p\}$. Appendix~G of Ref.~\cite{Komorovsky2016} evaluates Eq.~\eqref{eq:secondq} with the numbers $K_{qp}$ of Section~\ref{sec:sum} in place of $u_{qp}$, in a Kramers basis, $\phi_p=e_p$ and $\phi_{\bp}=\bar e_p$. By Eqs.~\eqref{eq:jB} and \eqref{eq:jBbar} and orthonormality, the array of these numbers is, in the ordering $(e_1,\dots,e_m,\bar e_1,\dots,\bar e_m)$ and with $\mathbf{1}$ the $m\times m$ unit matrix,
\begin{equation}
  \kappa=\begin{pmatrix}\mathbf{0}&-\mathbf{1}\\ \mathbf{1}&\mathbf{0}\end{pmatrix} ,
  \label{eq:Kelements}
\end{equation}
and substitution into Eq.~\eqref{eq:secondq} yields Eq.~(G2) of that work,
the operator
\begin{equation}
  A_B\equiv\sum_{p}\bigl(\ad{\bp}\an{p}-\ad{p}\an{\bp}\bigr) ,
  \label{eq:AB}
\end{equation}
which transfers, within each Kramers pair, an electron to its partner, with
opposite signs for the two directions. Each number $K_{qp}$ is well defined, and so is the operator \eqref{eq:AB}. Equation~\eqref{eq:secondq}, however, is a formula for sums of \emph{linear} operators, so evaluating it with the numbers $K_{qp}$ sums not the antilinear $\theta$ but the linear operator whose matrix the array $\kappa$ of Eq.~\eqref{eq:Kelements} is. By Eqs.~\eqref{eq:jB} and \eqref{eq:jBbar}, that operator is $j_B$. In coordinates, for $x=\sum_p c_p\phi_p$, the two readings of the one array are
\begin{equation}
  (\theta x)_q=\sum_p\kappa_{qp}\,c_p^{*} ,
  \label{eq:antiread}
\end{equation}
\begin{equation}
  (j_Bx)_q=\sum_p\kappa_{qp}\,c_p ,
  \label{eq:linread}
\end{equation}
and Eq.~\eqref{eq:secondq}, which operates on coefficients alone, takes the second. Hence $A_B=(j_B)_+$, the sum \eqref{eq:secondq} formed with $u=j_B$, and $A_B$ is the operator behind the $K_+$ matrix. Recording the antilinear $\theta$ on
a basis and extending linearly is the factorization \eqref{eq:split},
carried out implicitly, and since the numbers \eqref{eq:Kelements} were
obtained from $\theta e_p=\bar e_p$, the factor retained is the one
attached to the Kramers basis, not the one of the standard
representation. The published matrices are generated not from
Eq.~\eqref{eq:AB} but from Eq.~(78) of Ref.~\cite{Komorovsky2016},
$K_+^2=-N\hat 1+2\sum_{i<j}K_iK_j$. This makes no difference. Replacing each $K_i$ by $j_{B,i}$, the action of $j_B$ on electron $i$, and using $j_B^{2}=-1$ gives
$-N\hat 1+2\sum_{i<j}j_{B,i}j_{B,j}=A_B^{2}$ identically, so the
identification also covers the matrices as actually published. Everything Ref.~\cite{Komorovsky2016} asserts about $K_+^2$ is accordingly read here as an assertion about $A_B^2$.

The algebra to which $A_B$ belongs is a pseudospin. The single-particle
operators acting within the $p$th Kramers pair combine into
\begin{equation}
  t^{x}_p=\tfrac12\bigl(\ad{p}\an{\bp}+\ad{\bp}\an{p}\bigr) ,
  \label{eq:tx}
\end{equation}
\begin{equation}
  t^{y}_p=\tfrac{\mathrm{i}}{2}\bigl(\ad{\bp}\an{p}-\ad{p}\an{\bp}\bigr) ,
  \label{eq:ty}
\end{equation}
\begin{equation}
  t^{z}_p=\tfrac12\bigl(n_p-n_{\bp}\bigr) ,
  \label{eq:tz}
\end{equation}
which annihilate empty and doubly occupied pairs, act as
$\tfrac12\bm{\sigma}$ on a pair holding an unpaired electron, and obey
$[t^x_p,t^y_p]=\mathrm{i}t^z_p$ with cyclic permutations. This pseudospin belongs to the chosen pairing, not to a physical state, and is to be kept apart from the effective pseudospin~$\tfrac12$ that describes a physical Kramers doublet. Comparing
$t^y_p$ with Eq.~\eqref{eq:AB} and writing
$\mathbf{T}=\sum_{p=1}^{m}\mathbf{t}_p$ for the total pseudospin,
\begin{equation}
  A_B=-2\mathrm{i}\,T_y ,
  \label{eq:main}
\end{equation}
\begin{equation}
  A_B^{2}=-4\,T_y^{2} .
  \label{eq:mainsq}
\end{equation}
The $K_+^2$ matrix of Ref.~\cite{Komorovsky2016} is thus the matrix of
$-4T_y^{2}$, with the pseudospin $\mathbf{T}$ attached to the chosen Kramers
pairing. The $z$ component of the same algebra is in established use,
\begin{equation}
  T_z=\tfrac12\bigl(\hat N_{\text{unbarred}}-\hat N_{\text{barred}}\bigr) ,
  \label{eq:MK}
\end{equation}
whose eigenvalue on a determinant is the number $M_K$ by which determinants
are classified in Kramers-restricted configuration
interaction~\cite{Fleig2001,Fleig2012}. The single-particle operator whose
sum is $T_y$ is $\tfrac{\mathrm{i}}{2}j_B$, with eigenvalues $\pm\tfrac12$ and
eigenvectors $(e_p\pm\mathrm{i}\bar e_p)/\sqrt2$ within the $p$th pair.
These eigenvalues will be called weights, and the weight of a determinant
is the sum of the weights of its occupied spinors.

\subsection{Spectrum and eigenbasis}

The spectrum follows from the weights. Since $\mathbf{t}_p$ annihilates empty and full pairs and redistributes a single electron within a pair,
\begin{equation}
  \bigl[\mathbf{T},\,n_p+n_{\bp}\bigr]=0
  \qquad\text{for every pair $p$} ,
  \label{eq:pairocc}
\end{equation}
so the determinants can be grouped by their pair occupations (by which
pairs hold zero, one, or two electrons) and $A_B$ has no matrix elements
between different groups. Within a group in which $N_O$ pairs hold one
electron each, $T_y$ acts only on those $N_O$ electrons and is a component
of an angular momentum composed of $N_O$ pseudospins $\tfrac12$, with
eigenvalues $m_y$ of multiplicity $\binom{N_O}{N_O/2+m_y}$, so that
\begin{equation}
  \mathrm{spec}\,A_B^{2}
  =\Bigl\{\,-4m_y^{2}\ :\
  m_y=-\tfrac{N_O}{2},\,-\tfrac{N_O}{2}+1,\dots,\tfrac{N_O}{2}\,\Bigr\} .
  \label{eq:spectrumeq}
\end{equation}
The eigenvalue $-4m_y^{2}$ joins the weights $\pm m_y$, so its multiplicity is that of $m_y$, doubled for $m_y\neq0$. Writing these eigenvalues as $-k^{2}$ identifies the label of
Ref.~\cite{Komorovsky2016} as $k=2\lvert m_y\rvert$, twice the magnitude of
a pseudospin projection, with the values $k=N_O,\,N_O-2,\dots$, terminating
at $0$ for even $N_O$ and at $1$ for odd. The three eigenvalue statements
quoted at the head of this section follow, and so does every further property of the matrix listed in Table~\ref{tab:spectrum}, which collects them with their origins.

\begin{table}[h]
\centering
\caption{Reported properties of the $K_+^2$ matrix and their origin in the
identification $A_B=-2\mathrm{i}T_y$. Reality and symmetry of the matrix
are reported in Ref.~\cite{Komorovsky2016}; the eigenvalues and their
bounds are its Eqs.~(34) and (35); the binomial degeneracies are described
as ``Pascal's triangle'' behavior in Ref.~\cite{Bucinsky2016} and appear in Ref.~\cite{Komorovsky2016} in the eigenvalue multisets of the parity manifolds of its Appendix~I, which together carry them;
the matrix for two unpaired electrons is its Eq.~(84), diagonalized in its Eq.~(85); the inertness of
closed pairs is established in its Appendix~G; the diagonal is its
Eq.~(80), which credits Bu\v{c}insk\'y \emph{et al.}~\cite{Bucinsky2015};
the signed labels, equal to $2m_y$ here, are Eq.~(12) of Ref.~\cite{Gall2018}.}
\label{tab:spectrum}
\begin{tabular}{ll}
\toprule
reported property & origin \\
\midrule
matrix real and symmetric & $A_B$ real and antisymmetric in $B$ \\
eigenvalues $-k^2$, $k$ integral & $k=2\lvert m_y\rvert$ \\
$k\le N_O$ & $\lvert m_y\rvert\le N_O/2$ \\
$k\equiv N \pmod 2$ & $N_O\equiv N$ \\
binomial degeneracies & angular-momentum weights \\
two unpaired electrons: $-4,-4,0,0$
  & $m_y=-1,0,1$ with multiplicities $1,2,1$ \\
closed Kramers pairs contribute nothing & $\mathbf{t}_p=0$ on empty and full pairs \\
diagonal elements $-N_O$
  & $\langle T_y^{2}\rangle=N_O/4$ on a determinant \\
signed labels $k\in\{-N_O,-N_O+2,\dots,N_O\}$ & $2m_y$; the signs drop out of the square \\
\bottomrule
\end{tabular}
\end{table}

The eigenvectors follow as well. Because $A_B$ is a sum of single-particle operators, an eigenbasis can be assembled from determinants, pair by pair. Rotating every pair that holds an unpaired electron to the weight vectors
$(e_p\pm\mathrm{i}\bar e_p)/\sqrt2$ turns each such determinant into an
eigenvector of $A_B$, with $m_y$ equal to half the excess of $+$ over $-$
among those pairs, and the eigenspaces of $A_B^{2}$ are spanned by these
determinants together with their partners at $-m_y$. An eigenbasis of the same eigenspaces is obtained in Ref.~\cite{Gall2018} by numerically diagonalizing the matrix of $K_+^2$ over the $2^{N_O}$ determinants of barred and unbarred labels, split there into two parity blocks, which is not necessary (that work reports some $32$ gigabytes of memory for the eigenvectors at seventeen unpaired electrons), as the weight vectors can be written down directly.

\subsection{Generator relation and antilinear realization}

The identification also decides the generator relation, presented in Eqs.~(28) and (42) of Ref.~\cite{Komorovsky2016} as
\begin{equation}
  \Theta=\prod_{i=1}^{N}K_i=\prod_{i=1}^{N}\mathrm{e}^{\frac{\pi}{2}K_i}
   =\mathrm{e}^{\frac{\pi}{2}\sum_iK_i}=\mathrm{e}^{\frac{\pi}{2}K_+} ,
  \label{eq:generator}
\end{equation}
by which that work names $K_+$ the generator of the time-reversal operator, in analogy to the generation of a unitary operator by a Hermitian one. Exponentiating $A_B$ produces the unitary factor of time reversal. Since $j_B^{2}=-1$ gives
$j_B=\exp\bigl(\tfrac{\pi}{2}j_B\bigr)$, and since exponentiating a
single-particle sum applies the exponential to every occupied spinor,
\begin{equation}
  \exp\Bigl(\tfrac{\pi}{2}A_B\Bigr)\,(v_1\wedge\cdots\wedge v_N)
  =j_Bv_1\wedge\cdots\wedge j_Bv_N
  \;\equiv\;U_\Theta\,(v_1\wedge\cdots\wedge v_N) . \label{eq:generator2}
\end{equation}
The operator $U_\Theta$ so defined is unitary, represented in the determinant basis by a signed permutation matrix. It is not the time reversal $\Theta$ but its unitary factor. Lifting the factorization \eqref{eq:split} to $N$ electrons gives
\begin{equation}
  \Theta=U_\Theta\,\mathcal{C} ,
  \label{eq:manysplit}
\end{equation}
where $\mathcal{C}$ is the conjugation relative to the determinant basis
built from $B$, the map that leaves the determinants unchanged and
conjugates the coefficients of an expansion in them. $A_B$ generates the
unitary factor of time reversal, not time reversal.

The conjugation $\mathcal{C}$ also supplies the antilinear realization under which the expressions of Ref.~\cite{Komorovsky2016} become exact. The composition
\begin{equation}
  L_B \equiv A_B\,\mathcal{C} \label{eq:antilinear}
\end{equation}
is well defined and antilinear, a natural realization of the formal $K_+$ attached to the basis $B$. Reference~\cite{Komorovsky2016} itself records that $K_+$ is antilinear and not unitary [its Eq.~(27)]; $L_B$ has both properties, $L_B^{\dagger}L_B=-A_B^{2}\neq1$ for $2\le N\le2m-2$. On the tensor-product space (Appendix~\ref{app:antilinear}), its summands $j_{B,i}\mathcal{C}$, the rotation of electron $i$ combined with a conjugation of the entire state, satisfy $(j_{B,i}\mathcal{C})^2=-1$ and commute with one another, the two properties from which Ref.~\cite{Komorovsky2016} derives the eigenvalue statements, and
\begin{equation}
  L_B^{2} = A_B^{2}
  \label{eq:antisq}
\end{equation}
(Appendix~\ref{app:antilinear}). Under the substitution $K_+\to L_B$ the eigenvalue manipulations of Ref.~\cite{Komorovsky2016} go through and yield the matrix of Eq.~\eqref{eq:main}. $j_{B,i}\mathcal{C}$ is not the time reversal of electron $i$ (no operator acts so, by Section~\ref{sec:sum}), and the realization exists only relative to the chosen basis, through $\mathcal{C}$. It has the same form as the reading examined in
Section~\ref{sec:sum}, a sum of rotations times one global conjugation;
the two differ only in the basis to which the conjugation refers, the standard representation there, the Kramers basis here. No realization without such a choice exists. Any antilinear map is a linear map times a conjugation relative to a chosen basis, and a conjugation confined to
one electron is excluded by Section~\ref{sec:sum}, so every well-defined
antilinear realization of the sum contains a single global conjugation,
and the one attached to the basis of the construction leads to the square $A_B^{2}$ of Eq.~\eqref{eq:mainsq}. The apparent additivity whose interpretation Gall \emph{et al.}\ name as open resides in the notation, not in an operator -- the additive part is the linear factor $A_B$, and the conjugation enters once, globally.

The generator relation acquires an exact status under the same substitution. With the exponential series read on the underlying real vector space (Appendix~\ref{app:antilinear}),
\begin{equation}
  \mathrm{e}^{\frac{\pi}{2}L_B} =\begin{cases}\Theta, & N\ \text{odd},\\[2pt] U_\Theta, & N\ \text{even}.\end{cases} \label{eq:genparity}
\end{equation}
For odd $N$ the generator relation of Ref.~\cite{Komorovsky2016} is exact under this realization, and for even $N$ the exponential produces the unitary factor in place of the time reversal. The parity is not a defect of the realization. For any antilinear $L$, the relation $L\mathrm{i}=-\mathrm{i}L$ gives $\mathrm{e}^{tL}\mathrm{i}=\mathrm{i}\,\mathrm{e}^{-tL}$ on the underlying real vector space, so an antilinear $\mathrm{e}^{tL}$ forces $\mathrm{e}^{2tL}=-1$, and $\mathrm{e}^{\frac{\pi}{2}L}=\Theta$ forces $\Theta^{2}=-1$ and with it odd $N$. No antilinear generator produces the time reversal of an even-electron system.

\section{What the label measures}
\label{sec:label}

Reference~\cite{Komorovsky2016} proposes the eigenvectors of the $K_+^2$
matrix as relativistic configuration state functions and compares them, in
its Sec.~VII, with the eigenfunctions of $\mathbf{S}^2$. A Cartesian component of an
angular momentum refers to an axis, and in $A_B=-2\mathrm{i}T_y$ that axis is supplied by the chosen Kramers basis. The label is $k=2\lvert m_y\rvert$: the eigenvalues of $A_B^2$ are $-4m_y^2$, so writing them as $-k^2$ makes $k$ twice the magnitude of the projection $m_y$.

The comparison is most transparent where spin is a good quantum number. In
a two-component space with real spatial orbitals, take $e_p=\varphi_p\alpha$;
then $K_0$ leaves $e_p$ unchanged, $\bar e_p=\theta e_p=\varphi_p\beta$, and
the Kramers pairs are the $\alpha$ and $\beta$ spin orbitals of each spatial orbital, restricted spin orbitals in the usual sense. In this basis, Eqs.~\eqref{eq:tx}--\eqref{eq:tz} reduce to the ordinary spin operators, $\mathbf{t}_p=\mathbf{s}_p$, so that
\begin{equation}
  \mathbf{T}=\mathbf{S} ,
  \label{eq:TeqS}
\end{equation}
\begin{equation}
  A_B^{2}=-4S_y^{2} ,
  \label{eq:TeqSsq}
\end{equation}
and $k=2\lvert m_y\rvert$ with $m_y$ the projection of the total spin on the axis selected by the basis. Table~\ref{tab:kcsf} evaluates the four eigenvectors for two unpaired electrons in the same basis. The sector $k=0$ contains the singlet together with the $m_y=0$ component of the triplet. The label is the analogue of $\lvert M_S\rvert$ and, like a spin projection, takes different values on the components of one and the same multiplet, the triplet of Table~\ref{tab:kcsf} carrying $k=2$, $2$, and $0$. Determinants
are labeled by the spinors they occupy, $\Psi_{1\bar2}$ occupying $e_1$ and
$\bar e_2$, and a pair holding an unpaired electron is written $\ket{\up}$
or $\ket{\dn}$ according as the electron occupies $e_p$ or $\bar e_p$.

\begin{table}[h]
\centering
\caption{The eigenvectors of the $K_+^2$ matrix for two unpaired electrons,
Eq.~(85) of Ref.~\cite{Komorovsky2016}, evaluated in the basis of restricted spin orbitals.}
\label{tab:kcsf}
\begin{tabular}{lccl}
\toprule
eigenvector & $k$ & $\langle \mathbf{S}^2\rangle$ & character \\
\midrule
$(\Psi_{12}-\Psi_{\bar1\bar2})/\sqrt2$   & 2 & 2 & triplet \\
$(\Psi_{1\bar2}+\Psi_{\bar12})/\sqrt2$   & 2 & 2 & triplet \\
$(\Psi_{12}+\Psi_{\bar1\bar2})/\sqrt2$   & 0 & 2 & triplet \\
$(\Psi_{1\bar2}-\Psi_{\bar12})/\sqrt2$   & 0 & 0 & singlet \\
\bottomrule
\end{tabular}
\end{table}

The original authors recorded the signature of a projection themselves. Reference~\cite{Komorovsky2016} labels the spin eigenfunctions by superscripts, $\Psi^{1,1}$ and $\Psi^{1,-1}$ for the triplet components with $M_S=\pm1$ and $\Psi^{15/4,1/2}$ and $\Psi^{15/4,3/2}$ for two quartet components. Recovering the configuration state functions requires combining
$\Psi^{1,1}$ with $\Psi^{1,-1}$ and, for three unpaired electrons,
$\Psi^{15/4,1/2}$ with $\Psi^{15/4,3/2}$ -- in the latter case, in the words
of that work, ``and hence break the $S_z$ symmetry''.
Reference~\cite{Bucinsky2016} states in its conclusions that the
eigenfunctions are ``physically not suited to label spin or angular
momentum states/terms'', and had observed that the degeneracies and
eigenvalues of the labels follow the pattern of $S_z$ while the total
spin itself is not resolved by them.
Equations~\eqref{eq:TeqS} and \eqref{eq:TeqSsq} account for all three
observations and name the axis that the analogy with $S_z$ leaves
open.

The construction is not basis-covariant, and this can be seen before any state is chosen. The numerical array $K_{qp}=\braket{\phi_q|\theta\phi_p}$ of an antilinear operator does not transform like the matrix of a linear one: under a change of orthonormal basis with coefficient matrix $C$,
\begin{equation}
  K^{B'}=C^{\dagger}K^{B}C^{*} ,
  \label{eq:translaw}
\end{equation}
the conjugation coming from the coefficients of the ket, whereas the matrix of a linear operator transforms as $C^{\dagger}MC$. In every Kramers basis the array \eqref{eq:Kelements} is therefore the same $\kappa$, as it must be, since $\phi_{\bp}=\theta\phi_p$ there by definition. Were $A_B$ a single operator expressed in different bases, its matrix in $B'$ would be $C^{\dagger}\kappa C$, which for the rephasing $C=\mathrm{diag}(\mathrm{e}^{\mathrm{i}\chi},\mathrm{e}^{-\mathrm{i}\chi})$ of one pair is $\bigl(\begin{smallmatrix}0&-\mathrm{e}^{-2\mathrm{i}\chi}\\ \mathrm{e}^{2\mathrm{i}\chi}&0\end{smallmatrix}\bigr)$; the construction assigns $\kappa$. The two laws agree only when $C$ is real; for every other Kramers-basis change (the rephasing just displayed among them) $A_{B'}$ is a different operator, and the passage from $B$ to $B'$ is not a change of coordinates. The commutation asserted in Ref.~\cite{Komorovsky2016}, although written for operators, is accordingly a statement about a Hamiltonian and a Kramers basis together.

For a spin-free Hamiltonian, any two spin quantization axes are carried into one another by a spin rotation, which is a symmetry of $H$, and commutation with one component of $\mathbf{S}$ therefore entails commutation with every rotated component. The transformations that carry one Kramers basis into another are not symmetries of $H$ in general, so whether $H$ commutes with $A_B^{2}$ is a question about the pair $(H,B)$. The eigenvalues of $A_B^{2}$ form the same multiset in every Kramers basis; the basis determines which state carries which of them.

The axis is supplied by the chosen Kramers basis, a choice that is not unique. Within a single pair, $e_p$ may be replaced by any unit combination $e_p'=\alpha e_p+\beta\bar e_p$, with numerical coefficients $\alpha$ and $\beta$, whose partner is then
$\bar e_p'=\theta e_p'=\alpha^{*}\bar e_p-\beta^{*}e_p$; in particular the
rephasing $e_p'=\mathrm{e}^{\mathrm{i}\chi_p}e_p$ gives
$\bar e_p'=\mathrm{e}^{-\mathrm{i}\chi_p}\bar e_p$. Across pairs, every unitary transformation that preserves the Kramers pairing yields another Kramers basis. All of these bases carry the same barred and unbarred labels, and none of them is physically distinguished.

Under the rephasing, the creation operators acquire
$\ad{p}\to\mathrm{e}^{\mathrm{i}\chi_p}\ad{p}$,
$\ad{\bp}\to\mathrm{e}^{-\mathrm{i}\chi_p}\ad{\bp}$, and
Eq.~\eqref{eq:AB} becomes
\begin{equation}
  A_{B'}=-2\mathrm{i}\sum_p
  \bigl(\cos 2\chi_p\,t^{y}_p+\sin 2\chi_p\,t^{x}_p\bigr) . \label{eq:rephased} \end{equation} Each pair axis is rotated by $2\chi_p$ and is reversed at $\chi_p=\pi/2$. An example with two unpaired electrons shows how the eigenvalue changes under such a rephasing. For
$\ket{\psi}=(\ket{\up\up}-\ket{\dn\dn})/\sqrt2$, using
$t^y\ket{\up}=\tfrac{\mathrm{i}}{2}\ket{\dn}$ and
$t^y\ket{\dn}=-\tfrac{\mathrm{i}}{2}\ket{\up}$,
\begin{equation}
  T_y^{2}\,\ket{\psi}=\ket{\psi} ,
  \label{eq:kbefore}
\end{equation}
so that $k=2$, whereas reversing the first axis ($\chi_1=\pi/2$,
$\chi_2=0$) replaces $T_y$ by $T_y'=-t^y_1+t^y_2$ and
\begin{equation}
  T_y'\,\ket{\psi}=0 ,
  \label{eq:kafter}
\end{equation}
so that $k=0$. Neither the state nor $\theta$ has changed, and both bases are Kramers bases with the same barred and unbarred labels. Two aligned pseudospins have simply been relabeled as anti-aligned. The spectrum of the matrix is the same multiset in every Kramers basis -- the rephasing changes the operator, and with it the value assigned to the fixed state. Unequal rephasing is not required for this. A uniform $\chi_p=\pi/4$ replaces $T_y$ by $T_x$ in Eq.~\eqref{eq:rephased}, and $T_x\ket{\psi}=0$ for the same state, so that $k=0$ again. A uniform rephasing acts as the global phase of $\theta$, with $e_p\to\mathrm{e}^{\mathrm{i}\chi}e_p$ corresponding to $\theta\to\mathrm{e}^{-2\mathrm{i}\chi}\theta$. Reference~\cite{Komorovsky2016} fixes that phase to $-\mathrm{i}$, and the fixing does not remove the freedom -- for the fixed $\theta$, the rephased spinors again form a Kramers basis. The label changes under a choice that no phase convention removes.

The original authors point to this freedom themselves. Bu\v{c}insk\'y,
Mal\v{c}ek and Biskupi\v{c} describe it as ``the degree of freedom of the
rotation among spinors of a given Kramers pair (and/or the phase factor
value)'', note that it persists in the Kramers-restricted regime, and set
it aside on the ground that ``labels barred and unbarred
will be preserved when formally enforcing Kramers symmetry (even if the
phase factor of individual Kramers pairs would be different\ldots)''
[p.~1041 of Ref.~\cite{Bucinsky2016}]. The labels are indeed preserved, but Eqs.~\eqref{eq:kbefore} and \eqref{eq:kafter} show that the eigenvalue is not. A reported $k$ presupposes its basis, in the same way that a value of
$M_S$ presupposes a quantization axis.

Averaging over the axes isolates the part of the label that does not depend on the axis choices within a fixed decomposition into Kramers pairs. With the axis of each pair (every direction is reached by the pair mixing $e_p'=\alpha e_p+\beta\bar e_p$) distributed independently and uniformly over its unit sphere, the cross terms of $A_{B'}^{2}$ average to zero and
$\overline{(\mathbf{t}_p\cdot\hat{\mathbf{n}}_p)^{2}}=\mathbf{t}_p^{2}/3$, so that
\begin{equation}
  \overline{A_{B'}^{\,2}}=-\tfrac{4}{3}\sum_p\mathbf{t}_p^{2}=-\hat N_O ,
  \label{eq:average}
\end{equation}
the operator counting the unpaired electrons. For a fixed decomposition
into Kramers pairs, the content of the additive construction that does
not depend on the within-pair axes is the number of unpaired electrons,
the quantity that underlies its diagnostic
use~\cite{Bucinsky2015,KasperLi2020}. The decomposition itself remains a
choice, and $\hat N_O=\sum_p(n_p+n_{\bar p}-2n_pn_{\bar p})$ counts unpaired electrons relative to it. Because
the
eigenvalue presupposes its basis, the symmetry of the Hamiltonian does not
carry its conservation with it. The commutation asserted in Ref.~\cite{Komorovsky2016} is a separate
statement, and it is examined in the next section.

\section{Time-reversal invariance and the asserted commutation}
\label{sec:claim}

The proposed quantum number rests on the commutation relation asserted in
Eq.~(69) of Ref.~\cite{Komorovsky2016},
\begin{equation}
  \bigl[H^{\mathrm{DC}},\mathcal{K}_+^{2}\bigr]=0 .
  \label{eq:claim}
\end{equation}
Both sides are matrices, and $\mathcal{K}_+$ is its symbol for the sum. The Hamiltonian and the square of the $K_+$ matrix are represented on the space spanned by the Kramers-restricted
Slater determinants over $M$ four-spinors ($M=2m$ here), denoted $F(M,N)$ in Ref.~\cite{Komorovsky2016}, and that work formulates the assertion there
because, as it notes, commuting operators on a finite-dimensional space
share eigenvectors. By the reading of Section~\ref{sec:operator}, the
assertion is $[H,A_B^{2}]=0$, with $H$ the matrix of the Hamiltonian in
the determinant basis. The quantum number, the configuration-state-function proposal, and a degeneracy argument concluding that a level with $k\neq0$ is at least doubly degenerate all depend on it. However, the relation does not follow from time-reversal invariance, and
the derivation given for it is invalid.

Time-reversal invariance supplies a conjugated relation. A
single-particle Hamiltonian $h$ is time-reversal invariant when it
commutes with $\theta$, $h\theta=\theta h$. Writing $\theta=j_BC_B$ and
using that $C_BhC_B$ is the operator whose matrix in $B$ is the entrywise
conjugate $h^{*}$,
\begin{equation}
  h\,j_B=j_B\,h^{*} .
  \label{eq:tworeadings}
\end{equation}
For $N$ electrons the same step, applied to $H\Theta=\Theta H$ with the factorization \eqref{eq:manysplit}, gives
\begin{equation}
  H\,U_\Theta=U_\Theta\,H^{*} ,
  \label{eq:ThetaInv}
\end{equation}
with $H^{*}=\mathcal{C}H\mathcal{C}$ the entrywise conjugate of the
Hamiltonian matrix in the determinant basis. The symmetry thus lets the Hamiltonian commute with $j_B$ or $U_\Theta$ only up to a conjugation. For an invariant Hamiltonian, the plain commutation $[h,j_B]=0$ therefore holds exactly when the matrix of $h$ in $B$ is real, and with spin--orbit coupling the matrix elements have imaginary parts in general. Time-reversal invariance of $h$ and the plain commutation $[h,j_B]=0$ are thus two different conditions, and neither implies the other. $H=2T_y$ commutes with $U_\Theta$ without being time-reversal invariant, and the matrix of Eq.~\eqref{eq:witness} below is time-reversal invariant
without commuting.

Two Kramers pairs suffice to illustrate the difference. In the ordering
$(e_1,\bar e_1,e_2,\bar e_2)$, consider the Hermitian matrix
\begin{equation}
  h=\begin{pmatrix}\mathbf{0}&V\\V^{\dagger}&\mathbf{0}\end{pmatrix}
\label{eq:witness}
\end{equation}
with the coupling block $V=\mathrm{diag}(\mathrm{i},-\mathrm{i})$ between
the pairs. This matrix satisfies Eq.~\eqref{eq:tworeadings}, so it is time-reversal invariant. The commutator has norm $\Fnorm{[h,j_B]}=4$, with
$\lVert\cdot\rVert_{F}$ the Frobenius norm; in this ordering the matrix
of $j_B$ is block diagonal with the $2\times2$ blocks
$\bigl(\begin{smallmatrix}0&-1\\1&0\end{smallmatrix}\bigr)$. It
merely gives the coupling between the pairs an imaginary part, with the
partner element conjugated as Eq.~\eqref{eq:tworeadings} requires. For a single Kramers pair the difference does not arise, since Hermiticity
reduces Eq.~\eqref{eq:tworeadings} there to a real multiple of the
identity. For a single-particle $h$ alone, a basis with $[h,j_B]=0$
always exists, the Kramers eigenbasis of $h$. The matrix
\eqref{eq:witness} therefore establishes that invariance does not imply commutation. Whether a compliant basis exists for a full Hamiltonian is decided by the two-electron part.

Equation~\eqref{eq:claim} itself constrains matrix elements. Writing $P_\mu$ for the projector onto the eigenvalue $\mu$ of $T_y$ in the determinant space, so that $A_B^{2}=-4\sum_\mu\mu^{2}P_\mu$,
\begin{equation}
  P_\mu\bigl[H,A_B^{2}\bigr]P_\nu=-4\bigl(\nu^{2}-\mu^{2}\bigr)P_\mu HP_\nu ,
  \label{eq:blocks}
\end{equation}
so Eq.~\eqref{eq:claim} holds exactly when the Hamiltonian has no matrix element between eigenvectors of $T_y$ whose eigenvalues differ in magnitude,
\begin{equation}
  P_\mu HP_\nu=0 \quad\text{whenever}\quad \mu^{2}\neq\nu^{2} .
  \label{eq:blockcond}
\end{equation}
For a Hamiltonian $H=h_+$ built from a single-particle $h$ by Eq.~\eqref{eq:secondq}, and for $2\le N\le 2m-2$, this condition is equivalent to $[h,j_B]=0$; the proof is a counting argument over the determinant weights (Appendix~\ref{app:block}). Outside that range, $A_B^{2}$ is a multiple of the identity and Eq.~\eqref{eq:claim} holds trivially. The matrix of Eq.~\eqref{eq:witness} is time-reversal invariant and violates $[h,j_B]=0$, so it violates the block condition \eqref{eq:blockcond} on the $N=2$ sector, the range that the equivalence covers for $m=2$.

The two-electron part fails outright for an ordinary interaction. Take
two Kramers pairs of restricted spin orbitals, rephase the first pair by
$\chi_1=\pi/4$, leaving the second unchanged (an equally legitimate Kramers basis, Section~\ref{sec:label}), and let the interaction consist
of the single exchange integral $K_{\mathrm{ex}}$ between the two open
pairs. On the two-electron space of the unpaired electrons the Coulomb
interaction projects to
$(J_{12}-\tfrac12K_{\mathrm{ex}})-2K_{\mathrm{ex}}\,\mathbf{s}_1\!\cdot\mathbf{s}_2$,
with $J_{12}$ the Coulomb integral. The constant term drops out of every commutator, so the example is the Heisenberg exchange
$H_{\mathrm{ex}}=J_{\mathrm{ex}}\,\mathbf{s}_1\!\cdot\mathbf{s}_2$ with
$J_{\mathrm{ex}}=-2K_{\mathrm{ex}}$. In the rephased basis, Eq.~\eqref{eq:rephased} gives
\begin{equation}
  A_{B'}=-2\mathrm{i}\bigl(s^{x}_{1}+s^{y}_{2}\bigr) ,
  \label{eq:ABex}
\end{equation}
and direct evaluation yields
\begin{equation}
  \bigl[H_{\mathrm{ex}},\,A_{B'}^{2}\bigr]
  =2\mathrm{i}J_{\mathrm{ex}}\bigl(s^{z}_{1}-s^{z}_{2}\bigr)\neq0 ,
  \label{eq:Hexcomm}
\end{equation}
although $H_{\mathrm{ex}}$ is time-reversal invariant.

Restricted spin orbitals are four-spinors with vanishing small components, Kramers pairs under $-\mathrm{i}\Sigma_yK_0$ in the span of the kinetically balanced basis of the derivation, and Ref.~\cite{Komorovsky2016} extends the assertion itself to ``any approximate two-component Hamiltonians involving the Coulomb operator for electron-electron interaction'' [its Sec.~VI]. The projection discards nothing that could restore the commutation -- the projector $P$ onto the states with one electron in each open pair commutes with $A_{B'}^{2}$ by Eq.~\eqref{eq:pairocc}, so $P[V_C,A_{B'}^{2}]P=[PV_CP,A_{B'}^{2}]$, and Eq.~\eqref{eq:Hexcomm} is that block of the commutator with the full Coulomb interaction $V_C$; a single-particle part acts on the block as a scalar and cannot compensate. The interaction is where the question is decided -- for a single-particle part alone a compliant basis always exists, the Kramers eigenbasis of $h$, and only the two-electron integrals close that route.

The failure is not confined to the rephasing $\chi_1=\pi/4$. With a general rephasing angle, the same evaluation multiplies the right side of Eq.~\eqref{eq:Hexcomm} by $\sin2\chi_1$, which vanishes only where the rephased axis is again collinear with the common one. The Coulomb exchange thus fails to commute with the construction for all but special choices of the Kramers pairs, and the common-spin-axis basis of Section~\ref{sec:valid} is such a special choice.

Consequently, the asserted commutation does not follow from the symmetry of $H^{\mathrm{DC}}$, and for the Coulomb exchange it is false.

\section{Where the published derivation fails}
\label{sec:derivation}

The derivation given for Eq.~\eqref{eq:claim}, Appendix~E of Ref.~\cite{Komorovsky2016}, fails at two points: the two-electron identity on which it
rests is false already in its own operator-valued form, and the
concluding step exchanges antilinear matrix elements for the scalar
coefficients of second quantization. The appendix works with matrices in two bases, the orthonormal kinetically balanced atomic-orbital basis of the calculation~\cite{StantonHavriliak1984}, with four-spinor functions $X_\lambda$, and the molecular-orbital basis $\{\phi_q\}$, related by the molecular-orbital coefficient matrix of that work, written $C$ there and $C_{\mathrm{MO}}$ here to keep it apart from the conjugations $C_B$ and $\mathcal{C}$,
\begin{equation}
  \phi_q=\sum_{\lambda}X_{\lambda}\,(C_{\mathrm{MO}})_{\lambda q} .
  \label{eq:basischange}
\end{equation}
Its Eq.~(E5) gives the matrix of the time-reversal operator in the
atomic-orbital basis as
\begin{equation}
  K_{\lambda\tau}=\braket{X_\lambda|K|X_\tau} =-\mathrm{i}\Sigma_y K_0\,\delta_{\lambda\tau} ,
  \label{eq:E5}
\end{equation}
an operator-valued array. The conjugation $K_0$ is retained and conjugates every coefficient standing to its right. From it the appendix obtains the single-electron identity, its Eq.~(E13), for the matrix $D$ of
the single-electron Dirac Hamiltonian in the molecular-orbital basis,
\begin{equation}
  K_{pq}D_{qr}-D_{pq}K_{qr}=0 ,
  \label{eq:E13}
\end{equation}
with summation over the repeated index $q$, justified there by the unitarity of the molecular-orbital coefficients [its Eq.~(E12)]; and the two-electron identity, its Eq.~(E14),
\begin{equation}
  K_{pq}g_{qrst}-g_{pqst}K_{qr}+g_{srpq}K_{qt}-K_{pq}g_{srqt}=0 ,
  \label{eq:E14}
\end{equation}
with $g_{prst}=\iint r_{12}^{-1}\phi_p^{\dagger}(1)\phi_r(1) \phi_s^{\dagger}(2)\phi_t(2)\,dV_{12}$ the two-electron integrals, the index pairs $(p,r)$ and $(s,t)$ belonging to electrons 1 and 2, justified by the reality of the Coulomb interaction and by the commutation of the time-reversal operator with the basis, its Eq.~(E4). From the
two identities the appendix concludes, its Eq.~(E16), that
$[\hat H^{\mathrm{DC}},\hat{\mathcal{K}}_+]=0$, ``where \dots\ the
operators $\hat H^{\mathrm{DC}}$ and $\hat{\mathcal{K}}_+$ have the
standard form in the second quantization formalism''; the commutation
\eqref{eq:claim} would follow by squaring.

Read with the conjugation carried through, the single-electron identity is correct, and it is the invariance relation of
Section~\ref{sec:claim}. Under the basis change \eqref{eq:basischange}, the numerical arrays $K^{\mathrm{AO}}$ and $K^{\mathrm{MO}}$ of an antilinear operator, with elements $\braket{X_\lambda|\theta X_\mu}$ and $\braket{\phi_p|\theta\phi_q}$, are related by Eq.~\eqref{eq:translaw} with $C=C_{\mathrm{MO}}$. Sliding $K_0$
through the coefficients in Eq.~\eqref{eq:E13} places this conjugation
on $D$; with $K$ from here on the numerical array, the content of the
identity is
\begin{equation}
  K_{pq}D^{*}_{qr}-D_{pq}K_{qr}=0 ,
  \label{eq:E13conj}
\end{equation}
the matrix form of Eq.~\eqref{eq:tworeadings} and a correct statement.
The two-electron identity is not correct under any reading. For the two-electron integrals, time-reversal invariance supplies the Kramers relation
\begin{equation}
  g_{\bar p\bar r\bar s\bar t}=g_{prst}^{*}
  \label{eq:gkramers}
\end{equation}
for unbarred $p,r,s,t$: barring all four indices conjugates the integral, with the conjugation again on one side, as in
Eq.~\eqref{eq:tworeadings}. The two justifications quoted above establish relations of this kind -- the Kramers relations of the integrals -- and nothing further. Equation~\eqref{eq:E14} asserts more. Its four terms form two single-particle commutators, one on the index pair of electron 1 and one, in the exchanged writing of the integrals, on that of electron 2 -- formally the relation $[g,K_1+K_2]=0$, the invariance relation of the sum. Time-reversal invariance supplies the invariance relation of the product $\Theta=K_1K_2$ -- the product form of Section~\ref{sec:sum} -- both electrons reversed at once, and that is Eq.~\eqref{eq:gkramers}. Read this way, Eq.~\eqref{eq:E14} presupposes an electron-wise time reversal, which Section~\ref{sec:sum} excluded. Sliding the retained conjugation through the coefficients, as in Eq.~\eqref{eq:E13conj}, gives its content in terms of the numerical arrays,
\begin{equation}
  K_{pq}g^{*}_{qrst}-g_{pqst}K_{qr}+g_{srpq}K_{qt}-K_{pq}g^{*}_{srqt}=0 ,
  \label{eq:E14star}
\end{equation}
and reading the appendix without the conjugation removes the two stars. The identity is false in both forms. In the rephased basis of the counterexample of Section~\ref{sec:claim}, at the component $(p,r,s,t)=(1,1,2,\bar 2)$, the left side of Eq.~\eqref{eq:E14star} equals $(\mathrm{e}^{-2\mathrm{i}\chi_1}-1)K_{\mathrm{ex}}$, and without the stars it equals $(\mathrm{e}^{2\mathrm{i}\chi_1}-1)K_{\mathrm{ex}}$ -- $\sqrt2\,K_{\mathrm{ex}}$ in magnitude at $\chi_1=\pi/4$. Writing the two commutators without the exchange of the integrals moves the failure to the component $(1,2,2,\bar 1)$, with values $(1-\mathrm{e}^{2\mathrm{i}\chi_1})K_{\mathrm{ex}}$ and $(1-\mathrm{e}^{-2\mathrm{i}\chi_1})K_{\mathrm{ex}}$ of the same magnitude. The identity thus fails, in either form, for all but special rephasings, while the Kramers relation \eqref{eq:gkramers} holds for every choice of the Kramers pairs.

The second defect is the concluding step. The second-quantized operator invoked in its Eq.~(E16) is built by the single-particle formula \eqref{eq:secondq} (this is Appendix~G), and Section~\ref{sec:operator} identified the result as the linear $A_B$; the antilinear, operator-valued matrix elements are exchanged
for the scalar coefficients that the formula requires. For $[H,A_B]=0$,
the identities are needed with $K$ as a numerical array and without the
conjugation, and in that form the single-electron identity requires the matrix of $D$ in the Kramers basis to be real, the condition that Section~\ref{sec:claim} established for $h$ and that time-reversal invariance does not supply.

The appendix thus establishes neither its Eq.~(E16) nor the asserted commutation, and the conclusions resting on them (the quantum number, the status of the eigenvectors as symmetry-adapted configuration state functions, and the block diagonalization of the represented Hamiltonian by sectors of $k$) are unfounded. The eigenvectors themselves remain the orthonormal basis constructed in Section~\ref{sec:operator}. Only their status as symmetry-adapted functions of the Hamiltonian is lost.
The degeneracy argument of Eqs.~(72)--(74) of Ref.~\cite{Komorovsky2016}
falls with them. It pairs an eigenvector of the represented Hamiltonian with its image under the $K_+$ matrix, a partner that the asserted commutation would place at the same energy, and concludes that a level with $k\neq0$ is at least doubly degenerate -- but without the commutation, an eigenvector of the represented Hamiltonian need not be an eigenvector of $A_B^2$ and carries no value of $k$.

\section{What remains valid}
\label{sec:valid}

\subsection{The spin-free case}

If the Hamiltonian is spin free, $[H,\mathbf{S}]=0$, then in the basis of
restricted spin orbitals, a common spin axis for all pairs,
$\mathbf{T}=\mathbf{S}$ by Eq.~\eqref{eq:TeqS}, and $[H,A_B]=0$ follows
from spin symmetry, for a reason unconnected with time reversal.

The common axis carries that argument. Spin symmetry makes $H$ commute
with every component of the total spin, so a projection onto any one axis
shared by all pairs is conserved. Rephasing the pairs independently turns
their axes in different directions, and a sum
$\sum_p\mathbf{t}_p\cdot\hat{\mathbf{n}}_p$ with unequal
$\hat{\mathbf{n}}_p$ is no longer a component of the total spin. Spin symmetry then protects nothing, and Eq.~\eqref{eq:Hexcomm} shows the failure for an ordinary exchange interaction.

The spin-free case is where the eigenfunctions have been examined so far.
The algebraic work obtains them in the abstract determinant basis, where
no Hamiltonian enters, phenomenologically in
Refs.~\cite{Bucinsky2015,Bucinsky2016} and by diagonalization in Ref.~\cite{Gall2018}.

\subsection{The overlap diagnostic}

Reference~\cite{Bucinsky2015} evaluates the sum of squared overlaps
between the occupied spinors and their time reverses [its
Eqs.~(5)--(11)],
\begin{equation}
  D=\sum_{i,j=1}^{N}\bigl\lvert\braket{\theta\phi_i|\phi_j}\bigr\rvert^{2}
  =\sum_{i=1}^{N}\bigl\lVert P\,\theta\phi_i\bigr\rVert^{2} ,
  \label{eq:overlapdiag}
\end{equation}
with $P$ the projector onto the occupied space. The second form shows $D$ to be a property of that space alone, unchanged by any unitary mixing of the occupied spinors and independent of any pairing or choice of axes. Bounded by $N$, the sum satisfies, on a Kramers-restricted determinant,
\begin{equation}
  N-D=N_O ,
  \label{eq:overlapdeficit}
\end{equation}
because each doubly occupied pair contributes both of its spinors to $D$,
while a pair holding an unpaired electron contributes nothing, the time reverse of an unpaired spinor being unoccupied. When the occupied space is not closed under $\theta$, the difference $N-D$ need not be an integer. It measures how far that space is from its own time-reversed image, $N-D=\tfrac12\Fnorm{P-\theta P\theta^{-1}}^{2}$.

Kasper \emph{et al.}~\cite{KasperLi2020} use instead the expectation value
of the squared generator [their Eqs.~(20) and (21)]. For the neutral
sodium atom at the X2C Hartree--Fock level they report $-1.000286$ against
the ideal $-N_O=-1$, reduced to $-1.00003$ by state-averaged complete-active-space self-consistent field. On a single determinant the two diagnostics are one quantity (under an antilinear realization of the form \eqref{eq:antilinear}, formed on a basis containing the occupied spinors, the expectation of the square is $-N+D$, the value Ref.~\cite{Bucinsky2015} obtains in the Kramers-restricted limit [its Eqs.~(17)--(21)]), so the Hartree--Fock value records $D=9.999714$ for the eleven electrons of sodium. The two differ on multiconfigurational states, for which no single occupied space exists while the expectation value remains defined.

A value close to $-N_O$ is not evidence for the commutation. The diagonal elements of the $K_+^2$ matrix equal $-N_O$ on every Kramers-restricted determinant in every Kramers basis (Table~\ref{tab:spectrum}), so a wave function close to a single such determinant returns a value close to $-N_O$ irrespective of any commutation relation, and the deviation measures the Kramers contamination of the computed spinors, which is how Ref.~\cite{KasperLi2020} reads it. To our knowledge, none of Refs.~\cite{Bucinsky2015,Bucinsky2016,Gall2018,KasperLi2020} assigns $k$ to a computed eigenvector of a relativistic Hamiltonian or tests the asserted commutation.

\subsection{A test for the commutation}

Whether a Hamiltonian commutes with $A_B^2$ for a given choice of the Kramers pairs is decided by the block condition \eqref{eq:blockcond}. For a time-reversal-invariant single-particle Hamiltonian it can always be met, since the Kramers eigenbasis of $h$ conserves the weights. For an interacting Hamiltonian the two-electron integrals must conserve them as well. Since $A_B$ is a single-particle sum, this is checked directly from the integrals -- in the eigenbasis of $j_B$, in which every orbital carries a weight $w_p=\pm\tfrac12$,
\begin{equation}
  h_{pq}=0 \quad\text{unless}\quad w_p=w_q ,
  \label{eq:wtest1}
\end{equation}
\begin{equation}
  g_{prst}=0 \quad\text{unless}\quad w_p+w_s=w_r+w_t .
  \label{eq:wtest2}
\end{equation}
If both hold, then $[H,A_B]=0$ and Eq.~\eqref{eq:claim} follows a fortiori. The single-particle condition alone is not a substitute, since it is satisfied identically in the Kramers eigenbasis of $h$ and places no constraint on the two-electron integrals.

For an interacting Hamiltonian, one case is known in which the condition holds: the spin-free Hamiltonian in a basis of restricted spin orbitals, where spin symmetry enforces it and the sectors of $k$ duplicate the $\lvert M_S\rvert$ blocking. Already a single exchange integral between two pairs violates it for all but special choices of the Kramers pairs (Section~\ref{sec:claim}). No basis satisfying the condition for a Dirac--Coulomb Hamiltonian with spin--orbit coupling has been presented in Refs.~\cite{Bucinsky2015,Bucinsky2016,Komorovsky2016,Gall2018} or elsewhere.

\section{Conclusions}
\label{sec:concl}

This work has asked what the additive construction of
Refs.~\cite{Bucinsky2015,Bucinsky2016,Komorovsky2016} computes. The
answer is a single operator, and once it is identified the properties reported for the construction follow from the algebra of that operator; the quantum number claimed for it does not.

The $K_+$ matrix of Ref.~\cite{Komorovsky2016}, the array of time-reversal overlaps between the basis spinors inserted into the single-particle formula of second quantization, is the matrix of $A_B=-2\mathrm{i}T_y$, one Cartesian component of the Kramers pseudospin of the chosen pairing. Within the same algebra, the $z$ component is the projection number $M_K$; under the antilinear realization $L_B$ the generator relation of the proposal is exact for odd $N$ and yields the unitary factor of time reversal for even $N$. The reported spectrum, with its parity rule, its bound by the number of unpaired electrons and its degeneracies assembled from binomial weight multiplicities, is the weight structure of this angular momentum, and the eigenvectors follow in closed form, without numerical diagonalization. The algebra itself is not new: in the
quaternion formulation of relativistic self-consistent-field
theory~\cite{SaueJensen1999}, $j_B$ is the imaginary unit that
maps a spinor to its Kramers partner, and $T_z$ is the $M_K$ of
Kramers-restricted configuration
interaction~\cite{Fleig2001,Fleig2012}. New are the identification of
the published matrix as that of $-2\mathrm{i}T_y$, the axis dependence
of the label that follows from it, and the block condition for the asserted commutation.

The quantum number claimed for this construction does not stand. Time reversal of a single electron within a many-electron state admits no operator, so the proposed sum has no electron-wise meaning. Like a spin projection, the label $k=2\lvert m_y\rvert$ refers to an axis supplied by the basis and changes when the Kramers pairs are rephased, and the commutation asserted in Ref.~\cite{Komorovsky2016} does not follow from time-reversal invariance -- the Coulomb exchange between two Kramers pairs violates it, and Section~\ref{sec:label} explains why the commutation depends on the chosen pairing.

The diagnostic use remains meaningful. The overlap sum $D$ of Ref.~\cite{Bucinsky2015} depends on neither a pairing nor a choice of axes. On a Kramers-restricted determinant $N-D$ equals the number of unpaired electrons, Eq.~\eqref{eq:overlapdeficit}; when the occupied space is not closed under time reversal, $N-D$ measures how far it is from its own time-reversed image. For a fixed pairing, the
axis average \eqref{eq:average} retains the number of unpaired
electrons as well.
Sectors of $k$, by contrast, are not protected by time-reversal symmetry. The one known case in which the block condition holds is the spin-free Hamiltonian with a common spin axis, where the sectors duplicate the $\lvert M_S\rvert$ blocking; no basis meeting the condition for a Dirac--Coulomb Hamiltonian with spin--orbit coupling has been presented, and Section~\ref{sec:valid} gives the integral test by which such a claim could be established.

\section*{Acknowledgments}

\begin{sloppypar}
S.G.T. was supported by the Deutsche For\-schungs\-ge\-mein\-schaft (DFG)
under Project 535298924.
\end{sloppypar}

\appendix

\section{Properties of the antilinear realization \texorpdfstring{$L_B$}{LB}}
\label{app:antilinear}

The summands of $L_B$ act on the tensor-product space, where the conjugation is taken relative to the product basis built from $B$; on the antisymmetric subspace it agrees with $\mathcal{C}$. The matrix of $j_{B,i}$ in the product basis is real, so $(j_{B,i}\mathcal{C})^{2}=j_{B,i}j_{B,i}^{*}=j_{B,i}^{2}=-1$, and two such compositions commute because the rotations act on different factors while the conjugations cancel in pairs. The same reality gives Eq.~\eqref{eq:antisq}. With $A_B^{*}$ the entrywise conjugate of the real matrix of $A_B$ in the determinant basis,
\begin{equation}
  L_B^{2} = A_B\,\mathcal{C}A_B\,\mathcal{C}=A_BA_B^{*}=A_B^{2} .
  \label{eq:antisqproof}
\end{equation}

For the parity result \eqref{eq:genparity}, powers of $L_B$ alternate between $A_B^{2n}$ and $A_B^{2n}A_B\mathcal{C}$, so the exponential series sums to
\begin{equation}
  \mathrm{e}^{\frac{\pi}{2}L_B} =\cosh\Bigl(\tfrac{\pi}{2}A_B\Bigr) +\sinh\Bigl(\tfrac{\pi}{2}A_B\Bigr)\mathcal{C} .
  \label{eq:coshsinh}
\end{equation}
The eigenvalues of $\cosh\bigl(\tfrac{\pi}{2}A_B\bigr)$ are $\cos(\pi m_y)$, which vanish for half-integer $m_y$, and those of $\sinh\bigl(\tfrac{\pi}{2}A_B\bigr)$ are $-\mathrm{i}\sin(\pi m_y)$, which vanish for integer $m_y$. Since $N_O$ carries the parity of $N$, $m_y$ is half-integer exactly for odd $N$. For odd $N$ the cosh term vanishes and $U_\Theta=\exp\bigl(\tfrac{\pi}{2}A_B\bigr)$ reduces to $\sinh\bigl(\tfrac{\pi}{2}A_B\bigr)$, so the exponential of $L_B$ equals $U_\Theta\mathcal{C}=\Theta$; for even $N$ the sinh term vanishes and the exponential equals $U_\Theta$. This is Eq.~\eqref{eq:genparity}.

\section{Equivalence of the block condition and \texorpdfstring{$[h,j_B]=0$}{[h,jB]=0}}
\label{app:block}

Let $H=h_+$ and $2\le N\le 2m-2$. One direction is immediate. $[h,j_B]=0$ lifts to $[H,A_B]=0$, since the single-particle sum carries commutators to commutators, and the block condition \eqref{eq:blockcond} follows. For the converse, write $h=\bigl(\begin{smallmatrix}a&W\\W^{\dagger}&b\end{smallmatrix}\bigr)$ in the eigenbasis of $j_B$, with $a$ and $b$ the blocks within the weights $\pm\tfrac12$ and $W$ the block between them, so that
\begin{equation}
  [h,j_B]=\begin{pmatrix}0&2\mathrm{i}W\\-2\mathrm{i}W^{\dagger}&0\end{pmatrix} .
  \label{eq:hcomm}
\end{equation}
Each element of $W$ is a matrix element of $H$ between determinants whose weights differ by unity, which the block condition annihilates except between $\mu=\mp\tfrac12$ and $\pm\tfrac12$. That exemption cannot conceal an element when $2\le N\le 2m-2$. Consider an element $W_{uv}$, coupling an orbital $u$ of weight $+\tfrac12$ to an orbital $v$ of weight $-\tfrac12$, and a determinant that contains $v$ but not $u$, with $n_+$ and $n_-$ electrons on the two weights, so that $n_-\ge1$, $n_+\le m-1$, and the weight is $\mu=\tfrac12(n_+-n_-)$. The exempt case $\mu=-\tfrac12$ means $n_-=n_++1$, which parity excludes for even $N$; there a determinant with $n_+=n_-=N/2$ realizes the element. For odd $N$ the choice $n_+=(N+1)/2$ avoids it and respects both bounds, since $N\le2m-2$ then implies $n_+\le m-1$, and $n_-=(N-1)/2\ge1$ for $N\ge3$. The block condition therefore forces $W=0$, which is $[h,j_B]=0$.

\end{document}